# Eye tracking as a tool to quantify the effects of CAD display on radiologists' interpretation of chest radiographs

Daisuke Matsumoto, M.D.[a], Tomohiro Kikuchi, M.D., Ph.D., MPH.[a,b], Yusuke Takagi, M.Eng.[b], Soichiro Kojima, M.D.[a], Ryoma Kobayashi, M.D.[a], Daiju Ueda, M.D.,Ph.D.[b,c,d], Kohei Yamamoto, M.Sc.[a], Sho Kawabe, MBA.[b], Harushi Mori, M.D., Ph.D.[a]

a Department of Radiology, Jichi Medical University, 3311-1 Yakushi-ji, Shimotsukeshi, Tochigi 329-0498, Japan

b Medical AI Promotion Institute Co., Ltd., Life Science Building, 12-9 Nihonbashi Odenmacho, Chuo-ku, Tokyo, Japan

c Department of Artificial Intelligence, Graduate School of Medicine, Osaka Metropolitan University, Asahi-machi, Abeno-ku, Osaka 545-8585, Japan

d Center for Health Science Innovation, Osaka Metropolitan University, Asahi-machi, Abeno-ku, Osaka 545-8585, Japan

**Author Contributions:**

Daisuke Matsumoto: Writing - original draft, Tomohiro Kikuchi: Conceptualization, Methodology, Software, Formal analysis, Investigation, Data curation, Visualization, Supervision, Project administration, Writing - original draft, Yusuke Takagi: Methodology, Visualization, Writing - review & editing, Soichiro Kojima: Writing - review & editing, Ryoma Kobayashi: Writing - review & editing, Daiju Ueda: Writing - review & editing, Kohei Yamamoto: Writing - review & editing, Sho Kawabe: Writing - review & editing Harushi Mori: Writing - review & editing

**Funding:**

This research did not receive any specific grant from funding agencies in the public, commercial, or not-for-profit sectors.

**Abstract**

*Rationale and Objectives:* Computer-aided detection (CAD) systems for chest radiographs are widely used, and concurrent reader displays such as bounding-box (BB) highlights may influence interpretation. This pilot study used eye tracking to examine which aspects of visual search were affected.

*Materials and Methods:* We sampled 180 chest radiographs from the VinDR-CXR dataset: 120 with solitary pulmonary nodules or masses and 60 without. BBs were configured for 80% display sensitivity and specificity. Three radiologists (with 11, 5, and 1 years of experience) interpreted each case twice—once with BBs visible and once without—after a ≥2-week washout. Eye movements were recorded using an EyeTech VT3 Mini. Metrics included interpretation time, time to first fixation, lesion dwell time, total gaze-path length, and lung-field coverage. Outcomes were modeled using a linear mixed model with reading condition as a fixed effect and case and reader as random intercepts. Primary analysis was restricted to true positives (n=96).

*Results:* Concurrent BB display prolonged interpretation time by 4.9 s ($p<0.001$) and increased lesion dwell time by 1.3 s ($p<0.001$). Total gaze-path length rose by 2,076 pixels ($p<0.001$), and lung-field coverage increased by 10.5% ($p<0.001$). Time to first fixation was reduced by 1.3 s ($p<0.001$).

*Conclusion:* Eye tracking revealed measurable changes in search behavior associated with concurrent BB display during chest radiograph interpretation. These findings support this approach and highlight the need for larger studies across modalities and clinical contexts.

**Keywords:**

Eye-tracking; concurrent reader; chest radiography; bounding-box display; interpretation time

**Introduction**

Chest radiography (CXR) is one of the most widely performed imaging examinations worldwide and plays a critical role in the early detection of serious diseases, such as lung cancer and pulmonary infections[1]. Small pulmonary nodules are particularly prone to perceptual errors and missed diagnoses, highlighting the need for improved diagnostic accuracy[2].

Computer-aided detection/diagnosis (CAD) for CXR has emerged as a leading application of imaging artificial intelligence (AI), with nodule-detection systems now commercially available[3]. Among these, nodule detection CAD systems are widely available in many countries. Second-reader CAD preserves the opportunity for radiologists to perform habitual systematic searches before exposure to AI findings. In contrast, concurrent reader CAD may alter the search strategy from the beginning of interpretation, potentially changing how abnormalities are located and verified[4].

Previous studies have frequently reported changes in diagnostic accuracy and reading time with CAD/AI assistance[3,5,6]. By contrast, few studies have quantitatively assessed how CAD displays affect visual search behavior, including how radiologists initiate their search, verify detected abnormalities, and examine the image[7,8]. Eye tracking offers objective, fine-grained measurements of visual search processes, including time-to-first fixation, dwell time on the target, total gaze-path length, and coverage of the area of interest[7,9]. These metrics allow the quantitative assessment of the radiologists' visual search behavior under different CAD display conditions. Eye tracking research in radiology has characterized search strategies and error types[9]; however, its application in assessing the influence of different CAD display modes, especially concurrent CAD prompts, in CXR interpretation remains limited.

This pilot study aimed to explore whether eye tracking can serve as a useful tool for detecting differences in visual search behavior when radiologists interpret chest radiographs with concurrent-reader–style bounding-box highlighting. Understanding these effects may provide optimal strategies for AI-assisted CXR interpretation in clinical practice.

**Materials and Methods**

*Data Source and Preprocessing*

Images used in this study were drawn from the VinDR-CXR dataset[10]. Figure 1 illustrates the case selection flow. To ensure the feasibility of subsequent subgroup analyses, we secured at least 12 images in each group mentioned below, which corresponds to the minimum requirement for a paired t-test with Cohen's d = 0.8 (large effect size), one-sided α = 0.05, and 80% power. Based on these considerations, a total of 180 chest radiographs were ultimately included.

Among the selected images, 120 contained a solitary pulmonary nodule or mass and were randomly allocated into the following groups: (i) true-positive (TP), lesion-positive with a bounding box (BB) correctly marking the lesion location (n = 96); (ii) false-negative (FN), lesion-positive with no BB on the lesion and no incorrect BB elsewhere (n = 12); and (iii) FN plus false-positive (FP), lesion-positive with no BB on the lesion but with an incorrect BB elsewhere (n = 12). The remaining 60 lesion-negative images were randomly divided into (iv) true-negative (TN) with no BB (n = 48) and (v) FP images with a manually placed incorrect BB (n = 12). This allocation yielded an overall BB-display sensitivity of 96/120 (80%) and specificity of 48/60 (80%). For the TP cases, BBs were adopted from the VinDR-CXR ground-truth annotations, whereas for the FP cases, a radiologist manually placed BBs on structures commonly misinterpreted as pulmonary lesions, such as rib overlaps and nipple shadows.

### *Reader Study*

Two experimental conditions were tested: with and without CAD assistance. A crossover design was employed with each reader interpreting the same cases under both conditions. In session 1, the readers interpreted all cases under a simulated concurrent reader CAD setting; the BBs were displayed according to the predefined groups and placements described above. In session 2, after a washout period of at least 2 weeks, the same cases were re-read without CAD assistance, with the case order randomized for each session. Before the experiment, a separate set of 10 cases was used as a practice session to familiarize the readers with the graphical user interface.

Three radiologists (with 11, 5, and 1 year of experience) participated in the study. The readers were instructed to assess only lung fields. They were informed that the BBs represented AI outputs with nominal specifications of 80% sensitivity and 80% specificity; however, as described above, these BBs were manually created to achieve these figures. Readers were notified that eye-tracking would be performed, but the specific purpose was not disclosed.

### *Eye-Tracking Data Acquisition*

The eye movements were recorded using an EyeTech VT3 Mini device. The following metrics were extracted: (i) Total interpretation time; (ii) Dwell time on lesion: the cumulative duration during which the gaze remained within the lesion area, defined as the lesion BB plus a 50-pixel margin; (iii) Time to first fixation on lesion: the elapsed time from the start of case interpretation until the gaze first entered the defined lesion area; (iv) Total gaze-path length: the sum of gaze trajectory lengths measured in pixels; and (v) Lung-field coverage ratio—calculated as the proportion of 50-pixel lung-field grid cells crossed by the gaze path. Lung fields were generated using a deep-learning-based model (ChestXRayAnatomySegmentation; https://github.com/ConstantinSeibold/ChestXRayAnatomySegmentation). Additionally, a 50-pixel margin was applied to the gaze paths. A reading was considered valid if more than 50% of gaze samples were successfully captured during the interpretation period. Only valid readings were included in the subsequent analyses. Figure 2 shows an example of this visualization.

*Statistical Analysis*

Accuracy was calculated for each reader and each group, and the sensitivity and specificity were computed for each reader across all cases. The primary analysis was limited to the TP group (n = 96), and cases were included in the analysis only if eye tracking captured at least 50% of the gaze during the interpretation period. For each outcome metric, a linear mixed model (LMM) was fitted with the reading condition (session 1 vs. session 2) as a fixed effect, and the case and reader as random intercepts. Statistical significance was set at $p < 0.05$. The computational environment was Python 3.11 with the following key packages: Statsmodels 0.14.5, SciPy 1.16.1, NumPy 2.3.2, and Pandas 2.3.1.

As only publicly available data were used and no new information on the data was produced, Institutional Review Board approval was not required.

**Results**

In the TP group, valid experiments with sufficient gaze capture were obtained for 74 of the 96 cases for Reader A, 79 for Reader B, and 94 for Reader C.

Table 1 summarizes the interpretation accuracy and AI adoption rates stratified by the reader and case groups. In the TP group, the accuracy ranged from 63.5% to 82.3% in session 2 and from 83.3% to 99.0% in session 1, with all readers demonstrating improved accuracy when the BBs were displayed. Across all readers, sensitivity ranged from 60.0% to 79.2% and specificity from 63.3% to 96.7% in session 2, compared with a sensitivity of 79.2%–80.8% and specificity of 76.7%–86.7% in session 1.

Table 2 presents eye-tracking metrics by reader, along with overall comparisons derived from the LMM. Under the session 1 condition, interpretation time increased by a mean of 4.9 s (95% CI [3.3, 6.5], $p < 0.001$). Dwell time on the lesion increased by 1.3 s (95% CI [0.6, 1.9], $p < 0.001$). Time to first fixation on the lesion decreased by 1.3 s (95% CI [–2.1, –0.5], $p = 0.001$). Gaze-path length increased by 2,076 px (95% CI [1,337, 2,816], $p < 0.001$). Lung-field coverage ratio increased by 10.5% (95% CI [8.6, 12.4], $p < 0.001$).

**Discussion**

In this pilot reader study, eye tracking revealed measurable differences in interpretation time, lesion dwell time, time to first fixation, gaze-path length, and lung-field coverage under a concurrent BB display. These metrics provide a useful means for characterizing how CAD presentation modes influence radiologists' visual search behaviors during chest radiographic interpretation.

Previous studies have investigated the effects of second-reader CAD on diagnostic performance and reading time. However, the reported impact varies across modalities, diseases, and imaging settings, and a consistent picture has not emerged[11,12]. Because eye tracking offers the possibility of quantitatively assessing such CAD-related effects, recent studies have begun to apply it in this context. In a multicenter mammography study, Gommers et al. explicitly monitored readers' eye movements to compare performance and search patterns with and without AI decision support and reported significant differences in gaze behavior: fixation time within lesion regions increased, while overall coverage of the breast area decreased, alongside an improvement in diagnostic accuracy[8]. In the present study, using chest radiographs, interpretation time was prolonged under concurrent BB display, and gaze metrics shifted in a different manner from those reported by Gommers et al.[8]. Time to lesion localization was shortened, dwell time on the lesion increased, and overall lung-field coverage expanded. Taken together, the results indicate a shift toward a "find fast, verify thoroughly" reading pattern under concurrent prompting in the context of chest radiograph interpretation. These contrasts indicate that the influence of AI prompts on visual search behavior may differ according to modality and task, highlighting the importance of context-specific evaluation.

Attention to inter-reader variability is also important, as recent studies have shown heterogeneous impacts of AI-assistance across radiologists and tasks, and even experienced readers may remain susceptible to AI-driven influences such as automation bias[13,14]. In this study, the AI adoption rate (Table 1) and gaze metrics (Table 2) indicated inter-reader differences, suggesting differing degrees of reliance on BB's prompts. Thus, eye-tracking may provide a means to characterize reader-specific patterns of AI utilization. Although detailed inter-reader comparisons or typological classifications are beyond the scope of this pilot study, such analyses represent an important direction for future research and underscore the need for users to understand the specifications and performance characteristics of the products they deploy and to adapt their workflows accordingly.

As demonstrated in previous studies and this pilot investigation, the chosen presentation mode (e.g., second reader vs. concurrent reader) may influence radiologists' reading time and behavior. Therefore, it is desirable for product information to indicate how such modes could affect the reading process and how these effects are intended to be handled in practice. For products already in clinical use or those to be introduced in the near future, radiologists should be made aware of these potential effects, and CAD providers should bear the responsibility of communicating such information transparently. Recent societal statements also emphasize the importance of transparent information transfer from the provider to the deployer and clear user guidance regarding AI tools in radiology[15].

***Limitations and Future Directions***

This study had certain limitations. First, only three readers participated; thus, conclusions regarding the direction of inter-reader variability remain limited. Second, FP BBs were manually configured and therefore did not reflect the full error distribution or confidence information of the production AI systems. Third, the primary analysis was restricted to TP cases, and behavior under FP prompting was not evaluated (descriptive summaries of other case groups are provided in Appendix 1). Fourth, some cases were excluded because less than 50% of the gaze data were captured during interpretation; this likely reflects unfamiliarity with the experimental setup, although sufficient valid cases remained for analysis.

To address these limitations, we plan to expand the scale of the experiment and conduct it in the context of real-world AI use. We will also increase the number of readers to enable a more systematic evaluation of inter-reader variability and investigate the relationship between accuracy and eye-tracking metrics to gain further insight into how AI-assistance influences both performance and visual search behavior.

*Conclusion*

This pilot study demonstrated that eye tracking can capture how the presence or absence of CAD influences radiologists' interpretation of chest radiographs, identifying measurable changes across interpretation time, lesion dwell time, time to first fixation, gaze-path length, and lung-field coverage. These findings further suggest that reader responses to AI prompts vary across modalities, underscoring the need for broader context-specific investigations.

**List of abbreviations:**

AI:artificial intelligence, BB:bounding box, CAD: computer-aided detection, CI: confidence interval, CXR: chest x-ray, FN: false-negative, FP: false-positive, LMM: linear mixed model, TN: true-negative, TP: true-positive

**Funding:**

This research did not receive any specific grant from funding agencies in the public, commercial, or not-for-profit sectors.

**References**

1. Agrawal R, Mishra S, Strange CD, et al. The role of chest radiography in lung cancer. Semin Ultrasound CT MR 2024;45:430-439.

2. Gefter WB, Post BA, Hatabu H. Commonly missed findings on chest radiographs: causes and consequences. Chest 2023;163:650-661.

3. Bennani S, Regnard NE, Ventre J, et al. Using AI to improve radiologist performance in detection of abnormalities on chest radiographs. Radiology 2023;309:e230860.

4. Matsumoto S, Ohno Y, Aoki T, et al. Computer-aided detection of lung nodules on multidetector CT in concurrent-reader and second-reader modes: a comparative study. Eur J Radiol 2013;82:1332-1337.

5. Wenderott K, Krups J, Zaruchas F et al. Effects of artificial intelligence implementation on efficiency in medical imaging-a systematic literature review and meta-analysis. NPJ Digit Med 2024;7:265.

6. van Ginneken B, Schaefer-Prokop CM, Prokop M. Computer-aided diagnosis: how to move from the laboratory to the clinic. Radiology 2011;261:719-732.

7. van der Gijp A, Ravesloot CJ, Jarodzka H, et al. How visual search relates to visual diagnostic performance: a narrative systematic review of eye-tracking research in radiology. Adv Health Sci Educ Theory Pract 2017;22:765-787.

8. Gommers JJJ, Verboom SD, Duvivier KM, et al. Influence of AI decision support on radiologists' performance and visual search in screening mammography. Radiology 2025;316:e243688.

9. Bigolin Lanfredi R, Zhang M, Auffermann WF, et al. REFLACX, a dataset of reports and eye-tracking data for localization of abnormalities in chest x-rays. Sci Data 2022;9:350.

10. Nguyen HQ, Lam K, Le LT, et al. VinDr-CXR: an open dataset of chest X-rays with radiologist's annotations. Sci Data 2022;9:429.

11. Rodríguez-Ruiz A, Krupinski E, Mordang JJ, et al. Detection of breast cancer with mammography: effect of an artificial intelligence support system. Radiology 2019;290:305-314.

12. Peters AA, Wiescholek N, Müller M, et al. Impact of artificial intelligence assistance on pulmonary nodule detection and localization in chest CT: a comparative study among radiologists of varying experience levels. Sci Rep 2024;14:22447.

13. Dratsch T, Chen X, Rezazade Mehrizi M, et al. Automation bias in mammography: the impact of artificial intelligence BI-RADS suggestions on reader performance. Radiology 2023;307:e222176.

14. Yu F, Moehring A, Banerjee O et al. Heterogeneity and predictors of the effects of AI assistance on radiologists. Nat Med 2024;30:837-849.

15. Kotter E, D'Antonoli TA, Cuocolo R, et al. Guiding AI in radiology: ESR's recommendations for effective implementation of the European AI Act. Insights Imaging 2025;16:33.

**Figures**

**Figure 1. Case selection flows from the VinDR-CXR dataset**

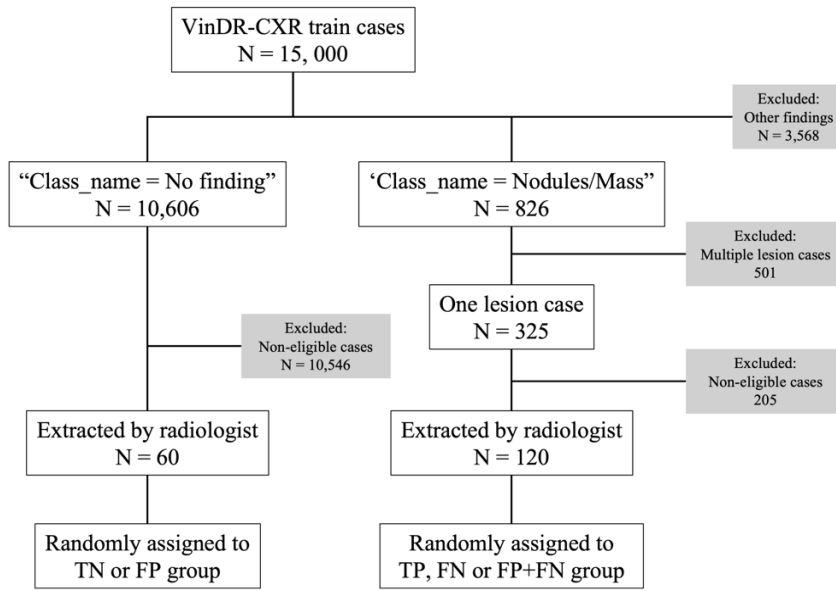

**Figure 2. Example of eye-tracking visualization**

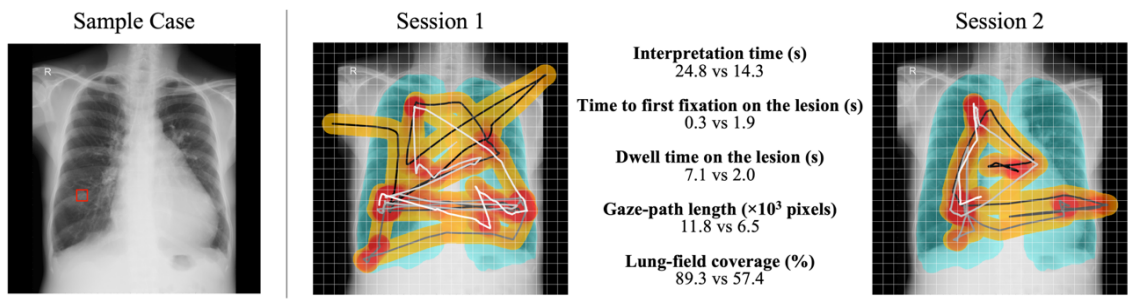

The cyan overlay masks the lung fields; the gaze path starts in black and ends in white, and a 50-pixel buffer around the path is shown as an orange band, whose overlap density is visualized as a heat map graduating from orange to red.

**Table 1. Diagnostic performance and AI adoption rates by reader and case group**

| Metric | Group | Radiologist 1 | | Radiologist 2 | | Radiologist 3 | |
|---|---|---|---|---|---|---|---|
| | | Session 1 | Session 2 | Session 1 | Session 2 | Session 1 | Session 2 |
| Accuracy (%) | TP | 83.3 | 69.8 | 99.0 | 82.3 | 85.4 | 63.5 |
| Improved_count | TP | 18 | | 17 | | 25 | |
| Worsened_count | TP | 5 | | 1 | | 4 | |
| AI adoption rate (%) | TP | 83.3 | | 99.0 | | 85.4 | |
| Accuracy (%) | FN | 66.7 | 58.3 | 16.7 | 66.7 | 58.3 | 41.7 |
| Improved_count | FN | 1 | | 0 | | 3 | |
| Worsened_count | FN | 0 | | 6 | | 1 | |
| AI adoption rate (%) | FN | 33.3 | | 75.0 | | 41.7 | |
| Accuracy (%) | FP+FN | 79.2 | 79.2 | 16.7 | 62.5 | 70.8 | 70.8 |
| Improved_count | FP+FN | 1 | | 0 | | 3 | |
| Worsened_count | FP+FN | 1 | | 11 | | 3 | |
| AI adoption rate (%) | FP+FN | 4.2 | | 33.3 | | 4.2 | |
| Accuracy (%) | FP | 91.7 | 100.0 | 16.7 | 58.3 | 83.3 | 100.0 |
| Improved_count | FP | 0 | | 0 | | 0 | |
| Worsened_count | FP | 1 | | 5 | | 2 | |
| AI adoption rate (%) | FP | 8.3 | | 66.7 | | 8.3 | |
| Accuracy (%) | TN | 72.9 | 93.8 | 97.9 | 64.6 | 87.5 | 95.8 |
| Improved_count | TN | 1 | | 16 | | 1 | |
| Worsened_count | TN | 11 | | 0 | | 5 | |
| AI adoption rate (%) | TN | 72.9 | | 97.9 | | 87.5 | |
| AI adoption rate (%) | Total | 71.4 | | 94.6 | | 77.4 | |
| Sensitivity (%) | Total | 80.8 | 67.5 | 80.8 | 79.2 | 79.2 | 60.0 |
| Specificity (%) | Total | 76.7 | 95.0 | 81.7 | 63.3 | 86.7 | 96.7 |

**Table 2.** Eye-tracking metrics by reader and overall comparisons based on linear mixed model analysis

| Metric | Radiologist 1 | | Radiologist 2 | | Radiologist 3 | | Diff (95% CI) | p |
|---|---|---|---|---|---|---|---|---|
| | Session 1 | Session 2 | Session 1 | Session 2 | Session 1 | Session 2 | | |
| valid cases (N) | 74 | 74 | 79 | 79 | 94 | 94 | | |
| Interpretation time (sec) | 14.6 [11.7, 22.1] | 10.2 [7.7, 13.2] | 5.2 [3.5, 7.9] | 4.4 [2.8, 7.6] | 13.2 [7.9, 25.6] | 8.6 [4.1, 15.2] | 4.9 (3.3 – 6.5) | < 0.001 |
| Dwell time on lesion (sec) | 2.2 [1.0, 3.7] | 1.2 [0.5, 2.1] | 1.1 [0.5, 2.1] | 0.4 [0.0, 0.9] | 3.3 [1.6, 6.7] | 1.3 [0.5, 3.7] | 1.3 (0.6 – 1.9) | < 0.001 |
| Time to first fixation (sec) | 1.0 [0.7, 2.4] | 1.9 [1.0, 7.0] | 1.8 [0.9, 3.5] | 2.7 [1.2, 5.0] | 0.7 [0.6, 1.4] | 1.4 [0.9, 3.8] | -1.3 (-2.1 – -0.5) | < 0.001 |
| Gaze path length (px) | 8459 [6334, 12272] | 5706 [4447, 9242] | 2764 [2028, 3547] | 2308 [1514, 3851] | 8383 [5692, 12180] | 4782 [3034, 7655] | 2076 (1337 – 2816) | < 0.001 |
| Lung-field coverage (%) | 85.5 [76.9, 88.5] | 68.0 [61.3, 77.7] | 52.9 [42.5, 62.6] | 46.8 [37.2, 56.7] | 73.9 [63.3, 81.2] | 62.4 [52.7, 72.7] | 10.5 (8.6 – 12.4) | < 0.001 |

Values are presented as medians [interquartile range, 25th–75th percentile].

**Supplemental Materials**

Appendix1

| Metric | group | Radiologist 1 | | | Radiologist 2 | | | Radiologist 3 | | |
|---|---|---|---|---|---|---|---|---|---|---|
| | | Session1 | Session2 | valid cases | Session1 | Session2 | valid cases | Session1 | Session2 | valid cases |
| **Interpretation time (sec)** | TP | 10.2 | 14.6 | 74 | 4.4 | 5.2 | 79 | 8.6 | 13.2 | 94 |
| | FN | 13.4 | 21.4 | 10 | 5.0 | 11.7 | 8 | 8.5 | 22.7 | 12 |
| | FP+FN | 13.0 | 21.6 | 21 | 9.8 | 10.7 | 18 | 9.3 | 20.9 | 23 |
| | FP | 12.2 | 21.6 | 11 | 11.7 | 9.3 | 10 | 10.1 | 20.9 | 11 |
| | TN | 10.4 | 17.6 | 35 | 16.8 | 4.1 | 40 | 9.9 | 18.5 | 48 |
| **Gaze path length (px)** | TP | 5706 | 8459 | 74 | 2308 | 2764 | 79 | 4782 | 8383 | 94 |
| | FN | 8012 | 8608 | 10 | 2155 | 4026 | 8 | 4395 | 10129 | 12 |
| | FP+FN | 7438 | 9399 | 21 | 3079 | 4128 | 18 | 5504 | 8575 | 23 |
| | FP | 7438 | 10097 | 11 | 4566 | 4409 | 10 | 6201 | 8575 | 11 |
| | TN | 7319 | 10160 | 35 | 4759 | 2230 | 40 | 5757 | 10511 | 48 |
| **Lung-field coverage (%)** | TP | 68.0 | 85.5 | 74 | 46.8 | 52.9 | 79 | 62.4 | 73.9 | 94 |
| | FN | 77.9 | 86.0 | 10 | 41.4 | 61.6 | 8 | 67.3 | 68.9 | 12 |
| | FP+FN | 72.4 | 87.1 | 21 | 51.9 | 59.3 | 18 | 70.7 | 75.2 | 23 |
| | FP | 65.1 | 88.2 | 11 | 56.8 | 56.9 | 10 | 75.1 | 75.7 | 11 |
| | TN | 74.8 | 84.1 | 35 | 58.3 | 50.5 | 40 | 72.8 | 84.2 | 48 |